\documentclass[11pt,prl,showpacs,preprintnumbers,floatfix,amsfonts,amssymb,amsmath]{revtex4}
\usepackage{graphicx}
\usepackage{bm}

\begin{document}

\title{Enhanced cooperativity below the caging temperature of \emph{o}-terphenyl} %

\author{B. M. Erwin}
 \affiliation{Department of Materials Science and Engineering and the Materials Research Institute\\%
  The Pennsylvania State University, University Park, PA 16802}
\author{S. Y. Kamath}
 \affiliation{Department of Materials Science and Engineering and the Materials Research Institute\\%
  The Pennsylvania State University, University Park, PA 16802}
\author{R. H. Colby}
 \affiliation{Department of Materials Science and Engineering and the Materials Research Institute\\%
  The Pennsylvania State University, University Park, PA 16802}

\date{\today}

\begin{abstract}
The utility of a cooperative length scale for describing the
dynamics of glass-forming liquids is shown using literature data on
\emph{o}-terphenyl. Molecular dynamics and Monte Carlo simulations
reveal a distribution of cooperative fractal events below the caging
temperature $T_A$. Guided by these results, we show how to extract
the size of slow regions in any glass-forming liquid from probe
rotation/diffusion measurements, which agrees quantitatively with
4-D NMR and grows steadily as temperature is lowered below $T_A$. We
clarify why this length must also be the size of the largest
cooperative events.
\end{abstract}


\maketitle

Many liquids either cannot crystallize or crystallize sufficiently
slowly that they vitrify below their glass transition temperature
$T_g$. A fundamental understanding of glass formation is still
lacking because it has not been firmly established whether the
pronounced slowing down is simply kinetic in origin or there is an
underlying thermodynamic character \cite{Adam1965, Kob1999,
Angell2000A, Sillescu1999, Ediger2000, Donth2001A}. Building upon
previous experimental results, this letter shows a natural length
scale for cooperative motion in \emph{o}-terphenyl that grows below
a \emph{caging temperature}, giving strong support to the
thermodynamic viewpoint \cite{Adam1965}. The caging temperature
$T_A$, is seen in many dynamics experiments on glass-forming liquids
\cite{Sillescu1999, Ediger2000, Angell2000A}. Particularly lucid are
experiments that measure translation and rotation of small probe
molecules immersed in the liquid. Above $T_A$, these are coupled by
the Stokes-Einstein relation and have the same temperature
dependence. In contrast, below $T_A$ these measurements decouple,
with translation having a weaker temperature dependence than
rotation \cite{Fujara1992, Hwang1996, Bainbridge1997, Hall1997,
Hall1998}. A wide variety of ideas have been expounded to explain
this observation \cite{Ediger2000, Donth2001A, Berthier2004B,
Schweizer2004A} and these ideas fall into two categories. One group
of ideas suggest that translation is dominated by a different and
faster moment of a distribution of relaxations than is measured by
rotation \cite{Ediger2000, Donth2001A}. The second group of ideas
consider both to be controlled by slow motions \cite{Berthier2004B,
Schweizer2004A}, and in this paper we present compelling evidence
that this is the correct picture.

The 1965 model of Adam and Gibbs \cite{Adam1965} suggests that there
should be cooperative motion in glass-forming liquids. The
cooperative size $\xi$, is inversely related to the configurational
entropy of the liquid $S_{conf}$. At temperatures sufficiently above
$T_g$, all molecules undergo independent local Brownian movements
without signs of cooperativity, where the relaxation time
$\tau_\alpha$, and the viscosity $\eta$, vary as
\begin{equation}\label{Eqn_high-T}
\tau_\alpha \sim \eta/T \sim \exp\left[E/k_BT\right] \qquad T>T_A
\end{equation}
where $E$ is the high-$T$ activation energy. As temperature is
lowered, the density of the liquid gradually increases and Brownian
motion becomes hindered, as neighboring molecules block each others
attempts to move. This crowding leads to \emph{cooperative dynamics}
\cite{Kob1999, Angell2000A, Kisliuk2000, Glotzer2000, Binder2003},
active for all $T$ below the \emph{caging temperature} $T_A$. The
onset of cooperativity is also accompanied by the observed `caging
effect' in the mean-square displacement of a particle between the
ballistic and diffusive regimes, and a reduction in $S_{conf}$,
causing $\xi$ to grow. These changes result in a progressively
stronger temperature dependence of $\eta$ and $\tau_{\alpha}$ at
lower temperatures.

Since experimental attempts to identify the length scale for
cooperative motion have met with limited success \cite{Sillescu1999,
Ediger2000, Donth2001A}, the dominant evidence for this quantity is
from computer simulations \cite{Glotzer2000, Binder2003,
Mountain1998, Donati1998, Donati1999A, Glotzer1999b, Muranaka2000b,
Kamath2002, Stanley2003, Lacevic2002}. Simulations have the profound
advantages of direct observation of motion and straightforward
identification of both the \emph{size and shape} of cooperatively
rearranged regions. Equilibrium simulations of liquids are not yet
possible near $T_g$, but have been done down to $0.7T_A$
\cite{Kamath2002}. In addition to confirming the essential aspects
of the Adam and Gibbs model, simulations have provided two novel
insights. Instead of a single size scale for cooperative motion,
there is in fact a \emph{broad distribution of size scales} below
$T_A$ \cite{Glotzer2000,  Mel'cuk1995, Mountain1998, Donati1998,
Donati1999A, Glotzer1999b, Muranaka2000b, Kamath2003, Stanley2003}.
The largest size  in this distribution $\xi$ grows rapidly as
temperature is lowered, as expected by Adam and Gibbs
\cite{Adam1965}. The second important observation is that the
cooperatively rearranging regions are not the three-dimensional
volumes that were initially proposed, but instead are \emph{fractal}
\cite{Glotzer2000, Donati1998, Donati1999A, Muranaka2000b,
Kamath2003}. The observed fractal dimension (of order $2$) clearly
shows that the majority of molecules within the volume $\xi^3$ have
\emph{not} participated in the cooperative event. Consequently, a
new model for cooperative motion was proposed that accommodates
these new insights \cite{Colby2000}. All glass-forming liquids show
a temperature dependence of cooperative size, with commensurate
effects in viscosity and relaxation time, in reasonable accord with
the expectations of dynamic scaling \cite{Colby2000, Erwin2002}.
\begin{equation}
\label{Eqn_xi} \xi \sim [(T-T_C)/T_C]^{-3/2} \qquad T_C < T < T_B
\end{equation}
\begin{equation}
\label{Eqn_tau} \tau_\alpha \sim \xi^6 \exp(E_\alpha/k_BT) \sim (T-T_C)^{-9} \exp(E_\alpha/k_BT) %
\end{equation}
\begin{equation}
\label{Eqn_eta} \eta \sim \xi^6 \exp(E_\eta/k_BT) \sim (T-T_C)^{-9} \exp(E_\eta/k_BT) %
\end{equation}
The critical temperature $T_C$, is \emph{below} $T_g$ and is the
temperature at which the \emph{equilibrium} extrapolated values of
$\xi$, $\tau_\alpha$ and $\eta$ all diverge. The material-specific
activation energy apparently depends on whether segmental relaxation
($E_\alpha$) or viscosity ($E_\eta$) is measured, with
$E_{\eta}/E_{\alpha}\simeq 1.2$ for non-polymeric organic glasses
\cite{Erwin2005A}. $T_B$ is the crossover temperature
\cite{Donth2001A, Beiner2001} below which all molecules are caged by
their neighbors ($T_g<T_B<T_A$). $T_B$ is typically about $20$ K
below $T_A$ and corresponds to the upper temperature limit where
dynamic scaling (and Vogel-Fulcher) quantitatively describe the
temperature dependence of relaxation.

\begin{figure}[t]
\centerline{\includegraphics[width=3.25in,keepaspectratio=true]{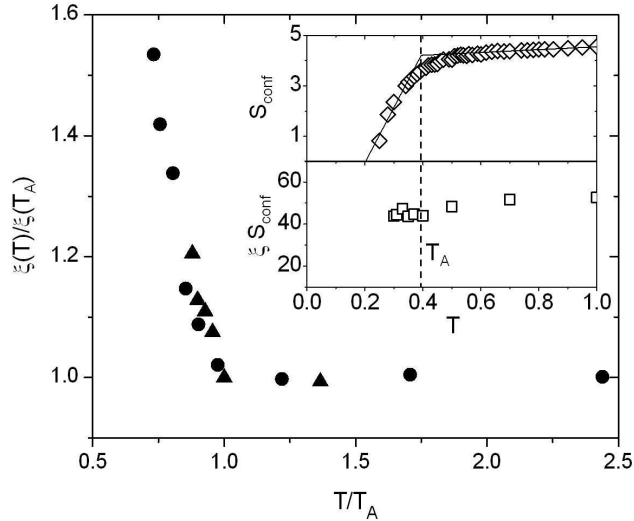}}
\caption{\label{Plot_simulation} Simulation results for the length
scale of cooperative motion as a function of reduced temperature.
From molecular dynamics simulations of Lennard-Jones sphere mixtures
($\blacktriangle$ \cite{Lacevic2002}) and from Monte Carlo
simulations of the bond fluctuation model at density $0.8$ for
chains of $10$ monomers ($\bullet$ \cite{Kamath2003}). Inset is the
temperature dependence of the configurational entropy ($\diamond$
\cite{Kamath2002}) and the product of cooperative length and
configurational entropy ($\square$ \cite{Kamath2003}) for chains of
10 monomers.}
\end{figure}

Figure~\ref{Plot_simulation} shows data for $\xi$ as obtained from
molecular dynamics simulations \cite{Lacevic2002} and new results
from Monte Carlo (MC) simulations on the bond fluctuation model
\cite{Kamath2002, Kamath2003}.
In the MC simulations the mobile particles are identified as those
that move over a time scale of interest. We then look for clusters
of these mobile particles and find that their sizes are a function
of time for short times, but quickly become time independent up to
$\tau_\alpha$. We only consider cluster sizes in this intermediate
time range. Figure \ref{Plot_simulation} shows the sizes of the
largest mobile particle clusters as a function of $T$. Both sets of
simulations show that $\xi$ is sensibly independent of temperature
above $T_A$, but then grows rapidly when temperature is lowered
below $T_A$. The very different nature of the simulations used for
the data in Fig.~\ref{Plot_simulation} suggests there may be
universal aspects to the growing length scale below $T_A$. The inset
of Fig.~\ref{Plot_simulation} shows that $S_{conf}$ as defined by
Adam and Gibbs also changes character at $T_A$ with a broad
crossover between $T_B$ \cite{Stickel1996} (below which
Eqs.~\ref{Eqn_tau} and \ref{Eqn_eta} or Vogel-Fulcher should
describe dynamics) and a much higher temperature above which
dynamics obey Eq.~\ref{Eqn_high-T} \cite{Stickel1996}. The inset
also shows that the product of $\xi S_{conf}$ is essentially
temperature independent. Hence, the growing length scale is
reciprocally related to the vanishing configurational entropy, in
qualitative agreement with Adam and Gibbs \cite{Adam1965}.

Guided by simulations, and owing to the abrupt change in the very
nature of relaxation at $T_A$ \cite{Binder1996, Baschnagel1997,
Zhang2000, Kamath2002, Binder2003}, the caging temperature is easily
identified by a variety of experiments probing liquid dynamics. We
demonstrate this point with \emph{o}-terphenyl using rotation and
translational diffusion of molecular probes, dielectric
spectroscopy, viscosity and Fabry-Perot interferometry. Molecular
probe techniques provide a \emph{model independent} measure of
$T_A$. The length scales extracted from these experiments, which are
in \emph{quantitative} agreement with existing $\xi(T)$ data from
4-D NMR \cite{Reinsberg2002, Qiu2002}, show a strong temperature
dependence \emph{only below} $T_A$. 4-D NMR \cite{Schmidt-Rohr1991,
Schmidt-Rohr1994, Heuer1995, Tracht1998, Reinsberg2001,
Reinsberg2002} provides the benchmark length scale of the slow
regions, but unfortunately this method can only be used over a very
limited temperature range \cite{Reinsberg2001}.

\begin{figure}[t]
\centerline{\includegraphics[width=3.25in,keepaspectratio=true]{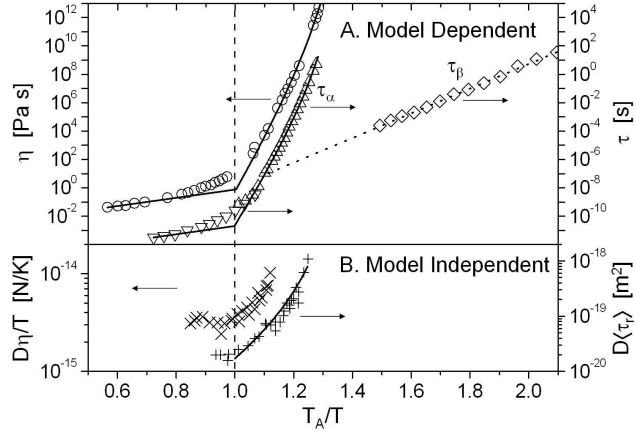}}
\caption{\label{Plot_Ta} Dynamic data for \emph{o}-terphenyl above
and below $T_A$. Model dependent measures of $T_A$ (A): viscosity
($\circ$ \cite{Greet1967, Laughlin1972}), dielectric spectroscopy
($\alpha$-relaxation ($\vartriangle$ \cite{Stickel1995}) and
$\beta$-relaxation ($\lozenge$ \cite{Wu1992})) and Fabry-Perot
interferometry ($\triangledown$ \cite{Steffen1992}). The dashed line
is $T_A$. The dotted line is a fit to the $\beta$-relaxation data
extrapolated to $T_B$. Solid curves are fits to $\tau_\alpha$ and
$\eta$ using Eq.~\ref{Eqn_high-T} for $T>T_A$ and Eqs.~\ref{Eqn_tau}
and \ref{Eqn_eta} for $T<T_B$. Model independent quantities affected
by $T_A$ (B): self-diffusion ($\times$ \cite{Fujara1992}) and
tetracene probe dynamics ($+$ \cite{Cicerone1995A, Cicerone1996}).
Solid curve is Eq.~\ref{Eqn_probe}.}
\end{figure}

$T_B$ was found to be $286$K from extrapolation of the
$\beta$-relaxation time to the $\alpha$-relaxation
(Fig.~\ref{Plot_Ta}.A), in agreement with the literature
\cite{Beiner2001}.
The temperature dependence of $\tau_\alpha$ and $\eta$
(Fig.~\ref{Plot_Ta}.A) were fit using Eqs.~\ref{Eqn_tau} and
\ref{Eqn_eta} and values for $T_C$, $E_\alpha$, and $E_\eta$ were
determined to be $227$ K, $120$ kJ/mol and $140$ kJ/mol
respectively. In figure \ref{Plot_Ta}.A, equation \ref{Eqn_high-T}
was fit to $\eta$ using $E = 22$ kJ/mol. This activation energy was
then used to fit $\tau$ (with a result that extrapolates to about
$10^{-14}$ sec at an infinite temperature). A $T_A$ of $313$ K was
then defined by the crossover between Eq.~\ref{Eqn_high-T} and
Eqs.~\ref{Eqn_tau} and \ref{Eqn_eta}. The $T_A$ from this crossover
between Eqs.~\ref{Eqn_tau} and \ref{Eqn_eta} and Arrhenius behavior
is also the temperature below which the Stokes-Einstein relation
breaks down. This `breakdown' below $T_A$ is apparent in figure
\ref{Plot_Ta}.B where both probe and self-diffusion show a
temperature dependence below $T_A$ \cite{Cicerone1995A, Fujara1992,
Cicerone1996}. The calorimetric $T_g$ of $o$-terphenyl was
previously determined to be at $241$ K \cite{Dixon1988}.

\begin{figure}[t]
\centerline{\includegraphics[width=3.25in,keepaspectratio=true]{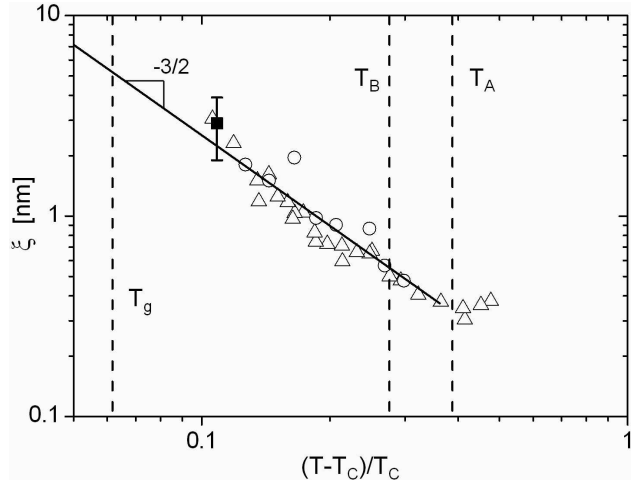}}
\caption{\label{Plot_xi} The temperature dependence of the
cooperative length scale in \emph{o}-terphenyl ($T_C = 227 $ K).
From 4-D NMR ($\blacksquare$ with error bars \cite{Qiu2002}) and
calculations using probe dynamics ($\circ$ anthracene
\cite{Cicerone1995A}; $\triangle$ tetracene \cite{Cicerone1995A}).
Dashed lines denote $T_g = 241 $ K, $T_B = 286 $ K and $T_A = 313 $
K. The solid line is the slope of $-3/2$ expected by dynamic scaling
(Eq.~\ref{Eqn_xi}).}
\end{figure}

The Stokes-Einstein relation expects the rotational relaxation time
$\langle\tau_r\rangle$ and translational diffusion coefficient $D$
of probe molecules are coupled so that their product is independent
of temperature. Above $T_A$ the probes diffuse a distance of order
their own size in the time it takes for the probe to rotate.
However, at $T_A$ these two dynamics \emph{decouple}, a fact which
can be used to establish the cooperative length scale. To understand
this, consider a bimodal distribution with a small quantity (volume
fraction $\phi_f << 1$) of fast and $\phi_s=1-\phi_f$ slow particles
\cite{Kumar2005}. The average rotational time is dominated by the
slow molecules
\begin{equation}
\langle\tau_r\rangle=\phi_s\tau_s+\phi_f\tau_f\cong\phi_s\tau_s
\end{equation}
since $\phi_s\gg\phi_f$ and $\tau_s\gg\tau_f$. The diffusion
coefficient is the sum of fast and slow contributions
\begin{equation}
D=\phi_sD_s+\phi_fD_f=\phi_s\xi^2_s/6\tau_s+\phi_f\xi^2_f/6\tau_f
\end{equation}
where $\xi_s^2\equiv6D_s\tau_s$ and $\xi_f^2\equiv6D_f\tau_f$. Since
$\phi_s\approx1$, the product
$6D\langle\tau_r\rangle\cong\phi^2_s\xi^2_s+\phi_f\phi_s\xi^2_f\tau_s/\tau_f$
can be used to define a length scale,
\begin{equation}\label{Eqn_probe}
\xi\cong\xi_s=\sqrt{6D\langle\tau_r\rangle} \textrm{.}
\end{equation}
Fig.~\ref{Plot_xi}.A shows the temperature dependence of this length
scale for two probe molecules in \emph{o}-terphenyl. We find that
the length scale defined by Eq.~\ref{Eqn_probe}
\emph{quantitatively} agrees with 4-D NMR, which is known to target
the slow contribution $\xi_s$.
This quantitative agreement requires $\phi_s^2\xi_s^2 \gg
\phi_f\phi_s\xi_f^2\tau_s/\tau_f$, justifying Eq.~\ref{Eqn_probe}.

In the calculation of $\xi$ over a wide temperature range,
$\langle\tau_r\rangle$ was interpolated using the noted relationship
$\langle\tau_r\rangle \sim \eta/T$ \cite{Bainbridge1997} along with
the viscosity data of Fig.~\ref{Plot_Ta}. $\xi$ from
rotation/translation of anthracene (circles in Fig.~\ref{Plot_xi})
shows a strong temperature dependence over the entire experimental
temperature range because all $T<T_A$, whereas the larger tetracene
probe (triangles in Fig.~\ref{Plot_xi}) gives $\xi$ that is clearly
temperature dependent below $T_A$ and independent of temperature
above $T_A$.

Ideally a probe molecule would have the same shape, size, polarity
and properties as the liquid matrix in which it was inserted.
$^1$H-NMR can provide measurements of translational and rotational
\emph{self}-diffusion coefficients \cite{Fujara1992}. This allows
for direct measurement of $\xi$ using Eq.~\ref{Eqn_probe}. Although
measurements of $\langle\tau_r\rangle$ exist \cite{Fujara1992}, they
are rare. In the case where measurements of $\langle\tau_r\rangle$
are lacking, we assume that $\eta/T$ properly describes the
temperature dependence of $\langle\tau_r\rangle$ \cite{Fujara1992,
Cicerone1996}. This assumption allows for measurements of $T_A$ that
stand in agreement with the other measurements of $T_A$ (see
$\times$ symbols in Fig.~\ref{Plot_Ta}), but with the limitation of
not being able to give an absolute measure of $\xi$.

In Fig.~\ref{Plot_xi}, measurements of $\xi$ are plotted along with
the absolute measurements of $\xi$ provided by 4-D NMR ($\xi = 2.9
\pm 1$ nm at $252 $ K \cite{Reinsberg2002}). While each technique
measures $\xi$ differently, they \emph{all stand in quantitative
agreement}. This result \emph{strongly suggests} that the vast
majority of the molecules are ``slow'', $\phi_f \approx 10^{-5}$,
which makes the slow term dominate even if $\tau_s/\tau_f=10^3$.
This concept of the slow molecules dominating near-$T_g$ dynamics
has also been suggested by recent models \cite{Schweizer2004A,
Berthier2004B}.

The cooperative length estimated from figure \ref{Plot_xi} is $5.2
\pm 1.0$ nm at $T_g$ and $0.33 \pm 0.06$ nm at $T_A$. The van der
Waals sphere radii $r_{vdW}$ of \emph{o}-terphenyl was calculated
from atomic radii using the procedures of Edward \cite{Edward1970}.
A $r_{vdW}= 0.37$ nm for \emph{o}-terphenyl is within the calculated
range of $\xi(T_A)$, suggesting that the magnitude of cooperative
size calculated herein is reasonable. This result $\xi(T_A) =
r_{vdW}$ appears to be general \cite{Erwin2005A}.

The length scale for cooperative motion has been estimated from
measurements of probe diffusion and rotation. At $T_g + 10$ K this
method agrees quantitatively with 4-D NMR and also provides a
measure of the cooperative size at higher temperatures.  This length
decreases as temperature is raised and adopts the van der Waals
radius of the glass-forming liquid above $T_A$.  The size of the
slow regions exhibits the temperature dependence expected by dynamic
scaling for the largest cooperatively moving (fast) regions,
suggesting that there is only one important length scale in this
problem. This makes sense because cooperative sizes need not grow
any larger than the size of the slow regions.

Despite the fact that the cooperative volume is fractal instead of
space-filling, the essential features of the Adam-Gibbs model
\cite{Adam1965} are correct. The cooperative size does indeed grow
rapidly as temperature is lowered below the caging temperature. Over
a temperature range extending from $T_A$ to $T_g$, $\xi$ of
\emph{o}-terphenyl has been observed to increase by an order of
magnitude from the van der Waals radius of each molecule, making the
insight of Adam and Gibbs particularly noteworthy.

We thank the National Science Foundation (DMR-0422079) for funding
and S.K. Kumar for discussions.

\end{document}